\documentclass{prop2015}% no class options needed by now
\usepackage[english]{babel}
\usepackage{picins}
\usepackage{graphicx}
\usepackage{subfigure}
\usepackage{hyperref}
\usepackage{bm}
\usepackage{float}
\usepackage[utf8]{inputenc}
\title{Analysis of universality in transient dynamics of coherent electronic transport}
\author[R. Seoane Souto]{R. Seoane Souto\inst{1}\footnote{Corresponding author\quad E-mail:~\textsf{ruben.seoane@uam.es}}}
\author[S.\,X. Author]{A. Mart\'{\i}n-Rodero\inst{1}}
\author[T.\,Y. Author]{A. Levy Yeyati\inst{1}}
\address[1]{Departamento de F\'{i}sica Te\'{o}rica de la Materia Condensada,\\
Condensed Matter Physics Center (IFIMAC) and Instituto Nicol\'{a}s Cabrera,
Universidad Aut\'{o}noma de Madrid E-28049 Madrid, Spain}
\shortauthors{R. Seoane Souto et al.}
\begin{abstract}
We analyze the time-dependent full-counting statistics of charges transmitted through a quantum dot in the coherent regime. The generating function for the time-dependent charge transfer statistics is 
evaluated
numerically by discretizing the Keldysh time contour, which allows us to compute the higher order charge and current cumulants. We also develop an analytic expression for all order cumulants at any 
time given as a function of the zeros of the generating function, finding that the short time \emph{universality} is due to the presence of a dominant single zero. The robustness of the universal features at short times
is studied in both the sequential and coherent regimes. 
\end{abstract}
\shortabstract
\begin{document}
\maketitle
%%% Use this if the article text won't start with a \section:
% \noindent
%%% Being based on LaTeX's article class, and2012 supports the respective 
%%% sectioning level from \section to \subparagraph.

\section{Introduction}
The analysis of the counting statistics, which concerns the probability distribution $P_q(t)$ of $q$ charge transfer events during a measurement time $t$, has attracted a lot of interest, since it 
provides information about interactions and correlations between electrons \cite{nazarov,blanter}. While traditionally these studies have been devoted to the stationary regime (long 
measuring times), there has been an increasing interest on the understanding of the time-dependent regime, triggered by recent advances on single electron sources \cite{feve,bocquillon,dubois} and the 
need to characterize these sources for detecting single electrons \cite{marquardt}.

In this context, some works have recently analyze the time-dependent transport statistics, studying the charge transferred cumulants \cite{PNAS_Flindt,Fricke,Fricke2} and the factorial cumulants \cite{Stegmann} in 
the incoherent regime. However, the coherent regime has been much less investigated \cite{Kambly}.  The existence of an universal scaling law for the higher order cumulants has been reported in both, the incoherent \cite{PNAS_Flindt} and the coherent \cite{DTA} regimes.  

The present work is devoted to the study of the full counting statistics in the time dependent regime, analyzing the origin and the robustness of the short time universal behavior. The system we consider is a spinless quantum dot (which will be referred to simply as \emph{dot} in what follows) coherently coupled to metallic electrodes. 
In order to study this system, we make use of Green function techniques, which are in principle the most appropriate tool to study the time-dependent properties for coherently coupled conductors both in the non-interacting \cite{Mukamel1,tang,tang2} and the interacting \cite{DTA} cases. 
We will compare two different parameter regimes corresponding to the incoherent (accessible by setting temperature much bigger than the tunneling rates coupling the dot and the metallic electrodes) and the coherent (zero temperature) situations. 
In the first part of the work we will focus on the uni-directional transport, considering that the voltage is the biggest energy scale in the system, while in  the final part we will relax this 
condition, and analyze the situation of bidirectional transport through the system.

Together with the numerical results, new simplified expressions are presented for the charge and the current cumulants as a function of the zeros of the generating function. These expressions are 
completely general, being valid not only in the transient regime, but also can be extended to the long time stationary case.
 
The paper is organized as follows: in Sect. \ref{Model} we introduce the theoretical model, together with the contour formalism and the simplified expressions for the charge and current characteristics.
In Sect. \ref{results} we analyze the results for the uni-directional case with one and two electrodes coupled to the dot, and the bidirectional one, showing their dependence with the zeros of the 
generating function. Finally, Sect. \ref{conclusions} is devoted to summarize the main conclusions of the work.

\section{Model and formalism}
\label{Model}
The system we consider is a quantum dot coupled to metallic electrodes, modeled by a single spinless level coupled to metallic electrodes. The Hamiltonian of the system 
is
\begin{equation}
 H=H_d+H_{leads}+H_T
\end{equation}
being $H_d=\epsilon d^{\dagger} d$, where $\epsilon$ is the bare electronic level and $d$ is the annihilation operator in the dot, 
$H_{leads}=\sum_{\nu k}\epsilon_{\nu k}c^{\dagger}_{\nu k} c_{\nu k}$  ($\nu =L,R$), where $\epsilon_{\nu k}$ are the leads electron energies, and $c^{\dagger}_{\nu k}$ are the corresponding creation 
operators acting on the electrodes. The bias voltage applied to the junction is imposed by shifting the chemical potentials of the electrodes $V=\mu_L-\mu_R$. The tunneling processes are described by 
\begin{equation}
 H_T=\theta(t)\sum_{\nu k}(\gamma_{\nu k}d^{\dagger} c_{\nu k}+\mbox{h.c.})\;,
 \label{tunnel}
\end{equation}
where $\gamma_{\nu k}$ are the tunneling amplitudes. Note that this part of the Hamiltonian has an explicit dependence on time through the Heaviside function. We define the tunneling rates 
$\Gamma_\nu=\mbox{Im}\sum_k|\gamma_{\nu k}|^2/(w-\epsilon_{\nu k}-i0^+)$, which can be considered constant in the so-called wide band approximation, and $\Gamma=\Gamma_L+\Gamma_R$.

In the present work we focus on the transient dynamics of the system from an initial $t=0$ configuration when the dot is suddenly connected to the electrodes
as described by Eq. (\ref{tunnel}). 
This situation might become experimentally accessible in the situation when the tunnel
 barriers can be controlled in times much smaller than the typical tunneling time for electrons. 

The corresponding properties can
be obtained form the Generating Function (GF), defined as
\begin{equation}
 Z(\chi,t)=\sum_{q}P_q(t)e^{iq\chi}\;,
\label{GF::prob}
\end{equation}
where the $P_p(t)$ denotes the probability of transferring $q$ electrons through the dot in the measuring time, $t$. The GF is related to the cumulant Generating function (CGF) by
$S(\chi,t)=\log\,Z(\chi,t)$ and all the charge transfer cumulants can be obtained from derivatives of this CGF as
\begin{equation}
 c_j(t)=\left.\left(\frac{\partial}{\partial i\chi}\right)^j S(\chi,t)\right|_{\chi=0}\;.
\label{def::c_j}
\end{equation}

\subsection{Keldysh contour integration}
\begin{figure}
     \begin{minipage}{1.0\linewidth}
      \includegraphics[width=0.9\textwidth]{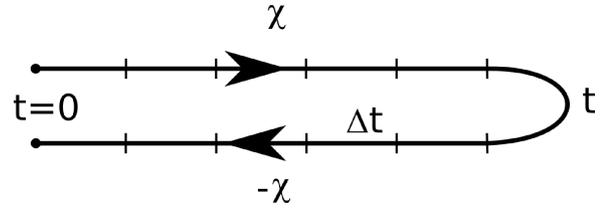}
     \end{minipage}
\caption{Keldysh contour considered to analyze the transient regime. $\chi$ indicates the counting field
changing sign on the two branches of the contour and $\Delta t$ corresponds to the time step in the
discretized calculation of the generating function $Z(\chi,t)$.}
\label{Keldysh-contour}
\end{figure}

The GF (\ref{GF::prob}) can be written \cite{Mukamel1} as an average of the evolution operator over the Keldysh contour, shown in Fig. \ref{Keldysh-contour}
\begin{equation}
 Z(\chi,t)=\left\langle T_c\exp\left\{-i\int_c \bar{H}_{T,\chi}(t')dt'\right\}\right\rangle_0 \;,
\end{equation}
where $\bar{H}_{T,\chi}$ is the tunnel Hamiltonian with a counting field $\chi(t)$ which takes the values $\pm\chi$ on the two branches of the Keldysh contour entering as a phase factor modulating the
tunnel amplitude, i.e.
\begin{equation}
 \bar{H}_{T,\chi}=\theta(t)\sum_{\nu k}\left(e^{i\chi_\nu}\gamma_{\nu k}c_{\nu k}^\dagger d+\mbox{h.c.}\right)\,.
\end{equation}

Notice that the different charge and current cumulants can be defined depending on how the phase $\chi(t)$ is distributed on the left and the right tunnel couplings. For instance, 
taking $\chi_L=\chi(t)$ and $\chi_R=0$, $Z(\chi,t)$ generates the current and charge transfer cumulants through the interface between the left lead and the dot. This is the choice that we shall
select for the rest of the work.

In Refs.\cite{utsumi,Mukamel1} it has been shown by path-integral methods that 
in the non-interacting case $Z(\chi,t)$ can be expressed as the following 
Fredholm determinant, defined on the Keldysh contour
\begin{equation}
Z(\chi,t) = \det\left(G \tilde{G}^{-1} \right)=
\det\left[ G\left(g^{-1}_0 - \tilde{\Sigma}\right)\right] \; ,
\label{z-nonint}
\end{equation}
where $\tilde{G}$ and $G$ denote the dot Keldysh Green functions,   
$g_0$ corresponds to the uncoupled dot case and 
$\tilde{\Sigma}$ are the self-energies due to the coupling to the leads.  
In the quantities $\tilde{G}$ and $\tilde{\Sigma}$ the $tilde$ indicates 
the inclusion of the counting field in the tunnel amplitudes.

As shown in \cite{kamenev} a simple discretized version of the inverse free dot Green 
function on the Keldysh contour is 

\begin{equation}
i g^{-1}_0 = \left(\begin{array}{cccc|cccc} -1 & & & & & & & -\rho \\
h_- & -1 & & & & & &  \\
& h_- & -1 & & & & &  \\
& & \ddots & \ddots & & & &  \\
\hline  
&  & & 1 & -1 & & &  \\
&  & & & h_+ & -1 & &   \\
&  &  &  & & \ddots & \ddots &  \\
&  &  & & & & h_+ & -1 \end{array} \right)_{2N\times2N} \; ,
\label{kamenev}
\end{equation}
where $h_{\pm} = 1 \mp i\epsilon_0 \Delta t$, $\Delta t$ indicates
the time step in the discretization with $N=t/\Delta t$. In this
expression $\rho$ determines the initial dot charge $n_d$ by $n_d = \rho/(1+\rho)$.

On the other hand, the dot self-energies are given by
\begin{equation}
\tilde{\Sigma}^{\alpha\beta}(t,t') = \alpha\beta \theta(t)\theta(t') \sum_{\nu k}
\gamma_{\nu k}^2 e^{i\left(\alpha-\beta\right)\chi_{\nu}/2}g^{\alpha\beta}_{\nu k}(t,t') \; ,
\label{non-interacting-sigma}
\end{equation} 
where $g^{\alpha\beta}_{\nu k}(t,t') = -i \langle T_{\cal C} c_{\nu k}(t_{\alpha}) c^{\dagger}_{\nu k}(t'_{\beta}) \rangle$, 
with $\alpha,\beta\equiv+,-$, are the Keldysh Green functions of the uncoupled leads. These self-energies can be evaluated on the time discretized time contour with $t_j=j\Delta t$
\begin{eqnarray}
 \Sigma^{+-}_{jk}&=&2i\sum_{\nu=L,R}e^{i\chi_\nu}\Gamma_\nu f^{\nu}_{jk}\nonumber\\
 \Sigma^{-+}_{jk}&=&2i\sum_{\nu=L,R}e^{i\chi_\nu}\Gamma_\nu\left[ f^{\nu}_{jk}-\delta[j-k]\right]\;,
\end{eqnarray}
where $f^{\nu}_{jk}$ are the Fourier transformed Fermi functions evaluated at a time $t=t_j-t_k$ and $\delta$ is the Kronecker delta function. The other two components are given by 
$\Sigma^{++}_{jk}=-\theta[j-k]\Sigma^{-+}_{jk}-\theta[k-j]\Sigma^{+-}_{jk}$ and $\Sigma^{--}_{jk}=-\theta[k-j]\Sigma^{-+}_{jk}-\theta[j-k]\Sigma^{+-}_{jk}$. The Fourier transformed Fermi function can be 
written as
\begin{eqnarray}
 f^{\nu}_{jk}&=&i\sum_{n=0}^{\infty}R_n\left[\theta[j-k]e^{\beta_n (j-k)\Delta t}\right.\nonumber\\
&&\left.-\theta[k-j]e^{-\beta_n (j-k)\Delta t}\right]e^{-i\mu_x (j-k)\Delta t}+\frac{\delta[j-k]}{2}, \nonumber\\ 
\end{eqnarray}
where $\beta_n$ and $R_n$ represent the poles and the residues of the Matsubara expansion, respectively, which depend on the temperature $T$. The convergence speed can be improved by using the 
approximated poles and residues proposed by T. Ozaki \cite{Ozaki} and computed using a continued fraction. The algorithm has been found to converge 
provided that $\Delta t\lesssim 1/(10\Gamma)$.

\subsection{Charge and current cumulants}
The expression of the GF (\ref{GF::prob}) can be seen as a generalized polynomial with real coefficients with $z=e^{i\chi}$ as the variable, i.e.
\begin{equation}
 Z(\chi,t)=z^{-q^-_{max}}\sum_{q=-q^-_{max}}^{q^+_{max}}P_q(t)z^{q+q_{min}}\;,
\label{GF::pol}
\end{equation}
where the sum has been truncated considering that until the measuring time $t$ there is a maximum amount of charge $q^+_{max}$ transferred from the (left) electrode to the dot, and the corresponding one in the opposite direction
$q^-_{max}$. The prefactor $z^{-q^-_{max}}$ has been taken outside the summation in order to avoid $1/z$ terms in the remaining polynomial. Another equivalent expression for the GF in terms of the zeros of the 
polynomial, $z=\alpha_k$ is
\begin{equation}
 Z(\chi,t)=z^{-q^-_{max}}\prod_{k=1}^N\left(z-\alpha_k\right)/P_{q^+_{max}}(t)\;,
\end{equation}
$N=q^+_{max}+q^-_{max}$ being the number of zeros of the GF. As 
shown in Refs. \cite{Utsumi,Kambly2,Abanov} in the absence interactions the zeros of the GF, or equivalently the singularities of the CGF,
are expected to appear in the real negative axis of the complex plane. Substituting this expression in Eq. (\ref{def::c_j}) we find
\begin{equation}
 c_j(t)=\left.\left(\frac{\partial}{\partial (i\chi)}\right)^j \left[\sum_k\left(e^{i\chi}-\alpha_k\right)+q^-_{max}\,e^{-i\chi}\right]\right|_{\chi=0}\;,
\label{cj_sum}
\end{equation}
which can be expressed as a sum of the contributions from the zeros of the GF, $\alpha_k$. This expression can be summed up exactly, finding
\begin{equation}
 c_j(t)=-\sum_k\mbox{Li}_{1-j}\left(\frac{1}{\alpha_k}\right)\;,
\label{charge::cum}
\end{equation}
where the $\mbox{Li}_{1-j}$ is the polylogarithm of order $1-j$ and we have made use of the symmetry with respect to the inversion of their argument $(-1)^n\mbox{Li}_{-n}(x)=\mbox{Li}_{-n}(1/x)$ (for 
$n>0$ and $x<0$).  Since $Li_{1-j}(0)=0$, the pole at $z=0$ due to the factorization in Eq. (\ref{GF::pol}) does not contribute to the transport properties. In Ref. \cite{DTA}, we
developed an approximate expression for Eq. (\ref{cj_sum}) in the case when the transport is dominated by a single process (the CGF is characterized by a single pole) and for orders $j\gg1$. Eq. (\ref{charge::cum}) 
constitutes a generalization of that expression for the case when more processes are involved and for any order $j$.

In the long-time regime, for the non-interacting case, the poles tend to accumulate on the negative real axis, defining two branch-cuts: one between $-\infty$ to a 
point $z^-$, and the other one between $z^+$ and $0$ \cite{Utsumi}, being the branch-points symmetrically located with respect to the point $z=e^{-\beta V/2}$. In Eq. (\ref{charge::cum}) the long time 
limit can be taken finding
\begin{equation}
\lim_{t\rightarrow\infty}  c_j(t) = -\int_{-\infty}^{z^-}dz \rho(z)\mbox{Li}_{1-j}\left(z\right)-\int_{z^+}^{0}dz \rho(z)\mbox{Li}_{1-j}\left(z\right)\;,
\end{equation}
where $\rho(z)$ is the distribution function for the poles. This stationary limit will be discussed elsewhere.

%If this distribution is uniform, the expression can be simplified by using the property $\int_{-\infty}^0 Li_{k}=0,\;\forall k\in\mathbb{N}^-$, as
%\begin{equation}
% c_j(t)=\int_{z^-}^{z^+}dz \mbox{Li}_{1-j}\left(z\right)\;.
%\end{equation}
%This stationary continuous distribution of poles will be discussed elsewhere.\newline

The current cumulants are defined as $I_j(t)=\frac{d}{dt}c_j(t)$, which recovers the zero-frequency steady state current cumulants when the measuring time $t\to\infty$. We can make use of the 
properties of the polylogarithms to determine the expression for the current cumulants
\begin{equation}
 I_j(t)=\frac{d}{dt}c_j=-\sum_k\frac{\alpha'_k}{\alpha_k}\mbox{Li}_{-j}\left(\frac{1}{\alpha_k}\right)\;,
\label{current::cum}
\end{equation}
which depends on the position, $\alpha_k$, and velocity of the pole drift, $\alpha'_k=\partial \alpha_k/\partial t\,$. 

%We can normalize these cumulants over the mean current, in order to avoid the dependency on the pole velocity, finding
%\begin{equation}
% \frac{I_j(t)}{I_1(t)}=(-1)^{j}\frac{\mbox{Li}_{-j}(\alpha)}{\mbox{Li}_{-1}(\alpha)}=\frac{\mbox{Li}_{-j}(1/\alpha)}{\mbox{Li}_{-1}(1/\alpha)}\;.
%\end{equation}
%This expression may diverge if the transport is not directional.\newline
\section{Results}
\label{results}
\subsection{Universal features in uni-directional transport}

\begin{figure}
\includegraphics[width=1\linewidth]{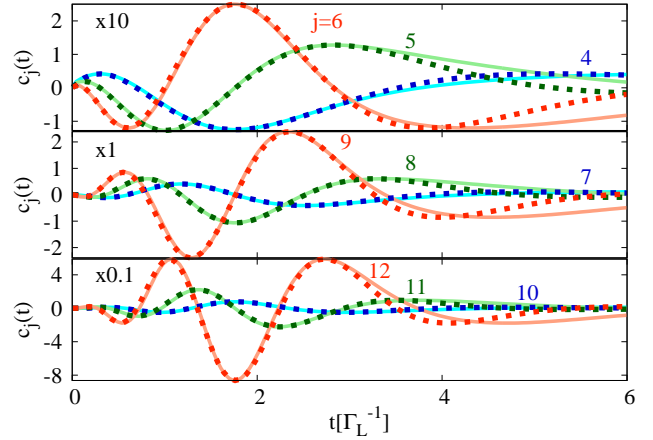}
\caption{(Color online): High order charge transfer cumulants comparing numerical results obtained by evaluating the Eq. (\ref{z-nonint}) (full lines) 
with the analytical expression of Eq. (\ref{charge::cum}) (dashed line) for the dot connected to only one electrode. The position of the pole is determined by Eq. \ref{dominant_pole}, with 
$\alpha_{long-t}=0$. The model parameters are $\Gamma_L=0.5$, $\Gamma_R=0$, $\mu_L=3$, $\epsilon=0$, $T=0.1$ and the dot is initially empty. (The same kind of oscillations are found in the sequential 
regime $\mu_L\gg k_bT\gg \Gamma$.)}
\label{fig1}
\end{figure}

We will first discuss the case where the transport in the short time regime trough the system is purely uni-directional. In order to force the system to exhibit an uni-directional transport, we consider 
the simple situation where a quantum dot prepared on a given initial configuration is suddenly coupled to only one single electrode. In this particular case 
and in the short time regime the GF can be approximated as 
\begin{equation}
 Z(\chi,t) \simeq P(0,t)+P(\pm1,t)e^{\pm i\chi}\;,
\end{equation}
where the sign $\pm$ denote the direction of the transport, determined by the initial population in the dot: positive for initially empty dot (transport from the electrode to the dot), and negative in the 
opposite case. Then, the charge cumulants can be obtained as derivatives of
the CGF as 
\begin{equation}
 c_j(t)=\left.\left(\frac{\partial}{\partial i\chi}\right)^j \log\left[1+\frac{1}{\alpha}e^{\pm i\chi}\right]\right|_{\chi=0}\;,
\end{equation}
where $\alpha=(-P(0,t)/P(\pm1,t))^{\pm1}$ is the only pole of the CGF. By using Eq. (\ref{charge::cum}), the charge cumulants have the simple expression
\begin{equation}
 c_j(t)=(-1)\mbox{Li}_{1-j}\left(\frac{1}{\alpha}\right)\;,
\label{charge_cum::1lead}
\end{equation}
where the $\mbox{Li }_{1-j}$ is the polylogarithm of order $1-j$. In Fig. \ref{fig1} we show the numerical results for high order cumulants of the transferred charge, compared with the analytical results.
The numerical results have been obtained by discretizing the Keldysh contour and evaluating numerically Eq. (\ref{z-nonint}). 
The time evolution of the pole of the CGF can be well approximated by
\begin{equation}
 \alpha_1\approx  -\frac{1}{(1-e^{-\Gamma_Lt})e^{2\Gamma_Lt}}+\alpha_{long-t}\;,
\label{dominant_pole}
\end{equation}
 with the only fitting parameter $\alpha_{long-t}$, the value of the pole at long times. The expression includes an exponentially decaying term to take into account the short time effects of the switching, and an exponentially increasing 
term to simulate the movement of the pole from $ -\infty$ to $0$ in the real negative axis. 

Notice that the analytic expression of Eq. (\ref{charge_cum::1lead}) is not only valid for the high order cumulants, but is also exact for any cumulant's order. As shown in Ref. \cite{DTA}, the high order charge cumulants exhibit an oscillatory behavior with amplitudes which scales as $\mbox{max}(c_j)\sim(j-1)!\pi^{(-j+1/2)}$.

\begin{figure}
\includegraphics[width=1\linewidth]{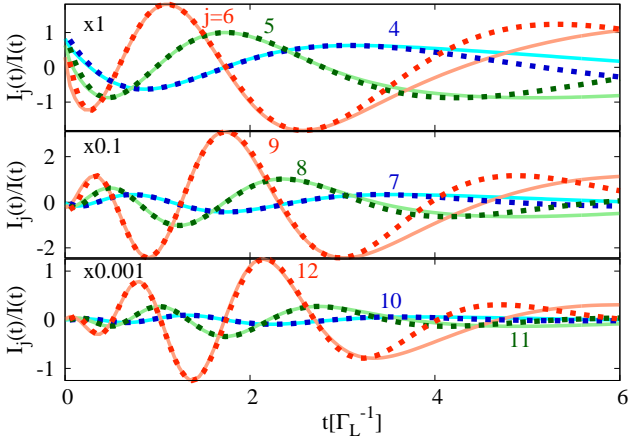}
\caption{(Color online): Current cumulants for the same case as in fig \ref{fig1} comparing the analytic results (discontinuous line) with the numerical ones (continuous).}
\label{fig_Icum}
\end{figure}

According to Eq. (\ref{current::cum}), the current cumulants have the simple expression
\begin{equation}
 I_j(t)=-\frac{\alpha'}{\alpha}\mbox{Li}_{-j}\left(\frac{1}{\alpha}\right)\;.
\end{equation}
This expression is dependent not only on the position of the pole, but also on the
velocity of its movement, breaking the universality at short times. In order to 
recover universality we can
normalize the current cumulants with the current, finding
\begin{equation}
 \frac{I_j(t)}{I_1(t)}=\frac{\mbox{Li}_{-j}(1/\alpha)}{\mbox{Li}_{-1}(1/\alpha)}\;,
\end{equation}
which, as the charge cumulants, exhibit an universal oscillatory behavior at short times (see Fig. \ref{fig_Icum}). As in the case of the charge cumulants, there is an universal scaling law for the current
cumulants $\mbox{max}(I_j/I_1)\sim j!\pi^{(-j-1/2)}$.
%This expression is, however, not complete since the poles at short times appear in pairs (they are degenerated). this fact leads to a factor $2$ in the expression of the cumulants, i.e
%\begin{equation}
% c_j(t)=2(-1)^{j}\mbox{Li}_{1-j}(\alpha)=2\mbox{Li}_{1-j}(1/\alpha)\;,
%\end{equation}
%Notice that the expression has not been normalized with the first cumulant of the charge transferred. The main reason for avoiding this normalization, is to avoid possible divergencies of the 
%quantities when the transport is not direccional ($c_1$ can be zero in this case). In any case (with or without the normalization), the higher order cumulants exhibit an universal short time oscillatory.

\subsection{Coherent effects in uni-directional transport}
\begin{figure}
\includegraphics[width=1\linewidth]{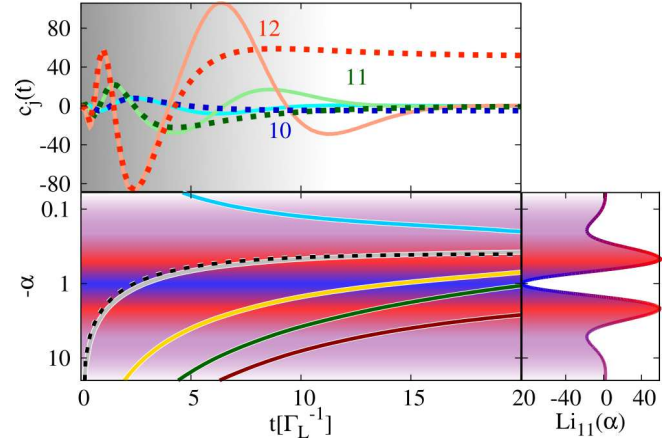}
\caption{(Color online): In the upper panel we represent the high order charge cumulants, $c_{10}$, $c_{11}$ and $c_{12}$ in the sequential regime comparing numeric results (continuous line) and the 
analytic one considering contributions from a single pole (discontinuous line) approximated by eq. (\ref{dominant_pole}) with $\alpha_{long-t}\approx-0.4$. In the lower panel we represent the time 
evolution of the dominant poles of the CGF and the approximated dominant pole (black discontinuous line),
 together with the polylogarithm $Li_{11}$ (related to the cumulant $c_{12}$). At short time, there is a single dominant pole (gray line) which leads to universality at short time (shadowed region in 
the upper panel), broken when more poles appear. The model parameters are $V=40$, $T=6$, $\epsilon=0$, $\Gamma_L=\Gamma_R=0.5$ and the dot is initially unpopulated.}
\label{sequential::Fig}
\end{figure}

In this section, we will discuss the the effect of attaching the system to a second electrode  simultaneously at $t=0$, in the regime where the transport is mainly uni-directional: $V\gg\Gamma$. Two
regimes are going to be considered: the \emph{sequential} regime ($V\gg T\gg\Gamma$) and the \emph{coherent} regime ($V\gg \Gamma\gg T$).\newline
We will first analyze the sequential case. In Fig. \ref{sequential::Fig} some of the high order charge cumulants are presented in the upper panel, comparing numerical results together with 
the analytical ones. For the analytic results we have considered that at short times the transport phenomena is dominated by a single pole given by Eq. (\ref{dominant_pole}).
At very short times, universal oscillations are observed (black shadowed region), which are 
signatures of uni-directional transport due to the dot charging process. At longer times ($t\sim 10\Gamma^{-1}_{L}$), the universality is broken due to the appearance of higher order processes, and the 
system evolves towards the stationary regime. A similar behavior is found by using the CGF provided by Flindt et. al. in Ref. \cite{PNAS_Flindt}.

In the lower panels of Fig. \ref{sequential::Fig} we show the first poles of the CGF (left panel) and the polylogarithm $Li_{11}$ (right panel), which is related to the charge cumulant $c_{12}$. The 
background color is used to indicate the height of $Li_{11}(\alpha)$. The dominant pole at short times (gray line), related to the dot charging process, appears at $z\to-\infty$ and 
evolves towards its stationary value, closer to $z=0$.
In its evolution, the pole crosses the region $z\sim1$, where the polylogarithms exhibit a strongly oscillatory behavior, producing the short time universal oscillations in the charge cumulants. 
At longer times, higher order processes become probable, leading to the appearance of more poles, breaking the universality and strong suppression the oscillations due to averaging over all the 
poles.\newline
As a final remark, in the lower panel of Fig. \ref{sequential::Fig}  there is a pole starting from $z=0$ and moving in the opposite direction than the dominant pole (blue curve).
This pole is caused by the bi-directionality of the transport through the system: due to the temperature there is a finite probability for the electron to be transferred in the opposite direction
than the bias voltage, due to the fact that T is not negligible compared to V ($V/T\sim6$). In the steady state, a gap at $e^{-\beta\mu_L/2}$ appears between the poles generated at $z\to-\infty$ 
and the ones at $z\to0$ \cite{Utsumi}.\newline

\begin{figure}
\includegraphics[width=1\linewidth]{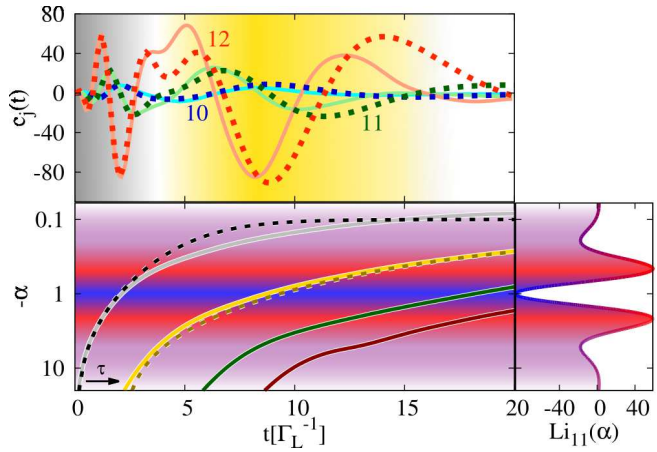}
\caption{(Color online): Upper panel: charge cumulants $c_{10}$, $c_{11}$ and $c_{12}$ comparing numeric results (continuous line) and the analytic one (discontinuous line).For the analytic result
two poles are considered, being the first one described by \ref{dominant_pole} with $\alpha_{long-t}=-0.1$, and the second one by eq. \ref{second_pole} with $\alpha_{2,long-t}=-0.15$.
At very short times, the universal oscillations are seen (black shadowed region), while at intermediate times a second family of (in general, non-universal) oscillations appear (yellow shadowed region). 
In the lower panels we represent the poles of the CGF (left) determined numerically (continuous lines) together with approximated (discontinuous lines), and the polylogarithm $\mbox{Li}_{11}(\alpha)$ 
(right). The model parameters are $V=6$, $T=0.1$, $\epsilon=0$, $\Gamma_L=\Gamma_R=0.5$ and the dot is initially unpopulated.}
\label{coherent::Fig}
\end{figure}

In the upper panel of Fig. \ref{coherent::Fig} we represent the same charge cumulants in the coherent regime and when the transport is still mainly uni-directional 
($V\gg\Gamma\gg T$). As in the incoherent case, at short times we observe the universal oscillatory behavior which is a signature of the dot charging process (black shadowed region). However, at 
intermediate times ($t\sim \Gamma/\Gamma_L\Gamma_R$) a secondary set of oscillations (yellow shadowed region) is observed. The shape and the amplitude of these second oscillations is not universal, since 
they depend, generally, on the model parameters.\newline
The existence of these two sets of oscillations can be understood by analyzing the pole's evolution of the CGF (lower panels of Fig. \ref{coherent::Fig}). Similarly to the sequential case, the universal 
oscillations appearing at short times are produced by the evolution of the dominant pole (gray line). The origin of the second set of oscillations is the appearance of a second pole (yellow curve) 
crossing the region of strong oscillations of the polylogarithm, approximated by
\begin{equation}
 \alpha_2\approx  -\frac{1}{(1-e^{-\tilde{\Gamma} (t-\tau)})e^{2\tilde{\Gamma}t}}+\alpha_{2,long-t}\;,\quad t>\tau\;;
\label{second_pole}
\end{equation}
with an effective rate $\tilde{\Gamma}=\Gamma_{L}^2\Gamma_R/\Gamma^2$, which takes into account that the second pole at short times is dominated by third order processes, and a delay time given by 
the characteristic time evolution of the first pole $\tau\approx 1/2\Gamma_L$. This second pole is related to a higher order process where the charge of the dot is relaxed through the right electrode, 
and the dot is charged again from the left one. The main difference with respect to the case studied before is the larger interval between the second pole and the next ones, avoiding them to interfere 
and producing the second set of oscillations. However, the poles are not totally independent and they can interfere, inducing a breaking on the universality on the second set of oscillations.

\subsection{Bidirectional transport}

\begin{figure}
\includegraphics[width=1\linewidth]{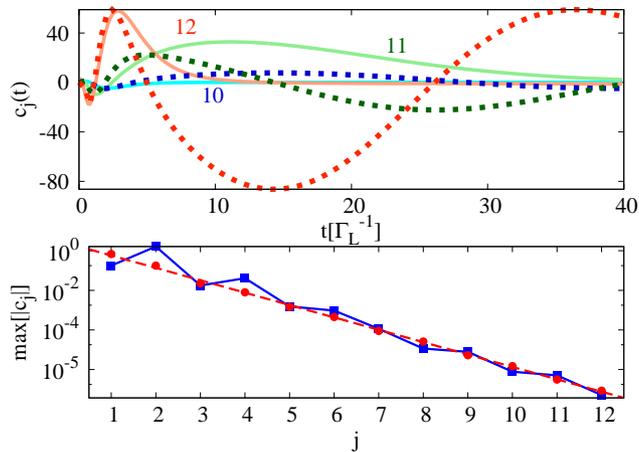}
\caption{(Color online): Upper panel, higher order charge cumulants (full line) showing their deviation with respect to the universal analytic result (dashed line) due to the 
bidirectional transport. The corresponding parameters are $V=0$, $\Gamma_L=\Gamma_R=0.5$, $\epsilon=0$, $T=0.1$ and the dot is initially empty. In the lower panel the amplitude of the oscillations are 
shown for the universal case (red dots), exhibiting an universal scaling of $\mbox{max}(c_j)\sim (j-1)!\pi^{(-j+1/2)}$, and for the bidirectional transport (blue squares), where the universal scaling is 
broken.}
\label{break_universality}
\end{figure}

In this section we will analyze the situation when the transport at short time is not uni-directional, but electrons are allowed to tunnel in both directions of the junction. This kind of situation
is found for an initially occupied dot or for $V\lesssim \Gamma$. In this section we will consider the limiting case when $V\approx 0$, where the effects are more pronounced.\newline
In the upper panel of Fig. \ref{break_universality} we show the higher order cumulants comparing the numerical results (full line) with the analytic ones corresponding to the case of a single pole is
involved in the transport, as described by Eq. (\ref{dominant_pole}) (dashed line). At very short times ($t\lesssim \Gamma^{-1}_L$) the dominant process corresponds to the dot charging and we observe 
again the universal features generated by the movement of a single dominant pole. However, at longer times, the universality is broken and we observe a new kind of oscillations with a renormalized 
amplitude. In the lower panel of Fig. \ref{break_universality} we represent the amplitude of the charge transfer cumulants oscillations with respect to the order, for the universal case (red dots) and the 
bidirectional one (blue squares). In the universal case the amplitude follows the scaling law $\mbox{max}(c_j)\sim (j-1)!\pi^{(-j+1/2)}$, while this law is broken in the bidirectional case.\newline
The breaking of the universality is due to the interference between two dominant poles: one starting at $z\to-\infty$, related to the dot charging from the left electrode, and another 
one at $z\to0$, related to the dot discharging through the left electrode. As the Fermi edges of the electrodes and the dot level are aligned, these two poles are equally dominant, since both processes 
are equally probable, leading to an interference that produces the universality breaking. The short time universality is recovered in the case when one of the two poles (or, equivalently, the transport 
in one of the directions) become more favorable, and one can estimate that this happens when the bias voltage becomes bigger than the tunneling rates ($V\approx\Gamma$).

%\subsection{local pairing in the dot?}
%The question here is to know if the universality can be also broken if the poles are splitted at very short times. One simple way to perform the test is to use a pairing potential localized in the dot,
%allowing interactions between spins. The main question here: Will this interation be \emph{fast} enough to split the poles at very short times?\\
%\emph{possible figures and comparisons}

\section{conclusions}
\label{conclusions}
In this work we have presented an analysis of the time dependent counting statistics of electron transport through a quantum dot coupled to metallic electrodes. We have focused on the analysis of the 
short time universality, developing new simplified analytic expressions as a functions of the poles of the CGF. We have understood that the universal oscillatory behavior of the 
higher order cumulants are generated by relaxation of the initial condition, which leads to a uni-directional transport. We have analyzed the sequential regime ($V\gg T\gg\Gamma$), where universal 
oscillations in the charge cumulants are observed at short time, and the coherent regime ($V\gg\Gamma\gg T$), where two sets of oscillations are found. Finally, we have analyzed the breaking of 
universality in the case of short time bidirectional transport occurring when the dot is on resonance and $V \lesssim \Gamma$.

\section{Acknowledgements}
The authors Acknowledge funding from MINECO through the grant FIS2014-55486-P. The authors thankfully acknowledges the computer resources, technical expertise and assistance provided by the Supercomputing 
and Visualization Center of Madrid (CeSViMa) and the Spanish Supercomputing Network (RES).

\end{document}